\documentclass[acmlarge]{acmart}

\usepackage{booktabs} % For formal tables
 \usepackage{url}

\usepackage[ruled]{algorithm2e} % For algorithms

\SetAlFnt{\small}
\SetAlCapFnt{\small}
\SetAlCapNameFnt{\small}
\SetAlCapHSkip{0pt}
\IncMargin{-\parindent}

\acmYear{2017}
\acmMonth{9}
\received{September 2017}
\begin{document}
% Title portion
\title{ProCMotive: Bringing Programability and Connectivity into Isolated Vehicles} 

\author{Arsalan Mosenia}
\affiliation{%
  \institution{Princeton University}
  \streetaddress{Department of Electrical Engineering, Princeton University}
  \city{Princeton}
  \state{NJ}
  \postcode{08540}
  \country{USA}}
\author{Jad F. Bechara}
\affiliation{%
  \institution{Princeton University}
  \streetaddress{Department of Electrical Engineering, Princeton University}
  \city{Princeton}
  \state{NJ}
  \postcode{08540}
  \country{USA}
}

\author{Tao Zhang}
\affiliation{%
  \institution{Cisco Systems}
  \streetaddress{}
  \city{}
  \state{}
  \postcode{}
  \country{USA}
}

\author{Prateek Mittal} 
\affiliation{%
  \institution{Princeton University}
  \streetaddress{Department of Electrical Engineering, Princeton University}
  \city{Princeton}
  \state{NJ}
  \postcode{08540}
  \country{USA}
}

\author{Mung Chiang}
\affiliation{%
  \institution{Purdue University}
  \streetaddress{Department of Electrical Engineering, Princeton University}
  \city{Princeton}
  \state{NJ}
  \postcode{08540}
  \country{USA}
}

\begin{abstract}
In recent years, numerous vehicular technologies, e.g., cruise control and steering assistant, have been proposed and deployed to improve the driving experience, passenger safety, and vehicle performance. Despite the existence of several novel vehicular applications in the literature, there still exists a significant gap between resources needed for a variety of vehicular (in particular, data-dominant, latency-sensitive, and computationally-heavy) applications and the capabilities of already-in-market vehicles. To address this gap, different smartphone-/Cloud-based approaches have been proposed that utilize the external computational/storage resources to enable new applications. However, their acceptance and application domain \textit{are still very limited due to programability,  wireless  connectivity, and performance limitations, along with several security/privacy concerns.}

In this paper, we present a novel reference architecture that can potentially enable rapid development of various vehicular applications while addressing shortcomings of smartphone-/Cloud-based approaches. \textit{The architecture is formed around a core component, called SmartCore, a privacy/security-friendly programmable dongle that brings general-purpose computational and storage resources to the vehicle and hosts in-vehicle applications}. Based on the proposed architecture, we develop an application development framework for vehicles, that we call \textit{ProCMotive}. ProCMotive enables developers to build customized vehicular applications along the Cloud-to-edge continuum, i.e., different functions of an application can be distributed across SmartCore, the user's personal devices, and the Cloud. 

In order to highlight potential benefits that the framework provides, we design and develop two different vehicular applications based on ProCMotive, namely, Amber Response and Insurance Monitor. We evaluate these applications using real-world data and compare them with state-of-the-art technologies. 
\end{abstract}

\keywords{Architecture, Application development, Cloud, Internet connectivity, Smartphone, Smart vehicles, Performance, Privacy, Programability, Security}

\thanks{

  Authors' addresses: A.~Mosenia, Department of Electrical Engineering, Princeton University, Princeton, NJ, 08540, US; J. F.~Bechara, Department of Electrical Engineering, Princeton University, Princeton, NJ, 08540, US; T.~Zhang, SJC09/2, 260 East Tasman Drive, San Jose, CA 95134; P.~Mittal, Department of Electrical Engineering, Princeton University, Princeton, NJ, 08540, US; M. Chiang, Electrical and Computer Engineering Department, Purdue University, West Lafayette, IN 47907.
}

\maketitle

\renewcommand{\shortauthors}{A. Mosenia et al.}

\section{Introduction}
\label{INTRO}
Rapid technological advances in sensing, signal processing, low-power electronics, and wireless networking are revolutionizing vehicle industry. To enhance the driving experience, passenger safety, and vehicle performance, numerous vehicular technologies have been suggested and partially deployed in recent years. For example, steering assistance and cruise control have been already integrated into state-of-the-art vehicles, and vision-based collision avoidance \cite{VBCA0,VBCA1} and sign detection \cite{SIGN_D1,SIGN_D2} have shown promising results and garnered ever-increasing attention from vehicle manufacturers. However, there still exists a significant gap between resources needed for a variety of vehicular (in particular, data-dominant, latency-sensitive, and computationally-heavy) applications and the capabilities of already-in-market vehicles \cite{GAP1, GAP2}.

A few vehicle manufacturers (for example, Tesla and Toyota) and several third-party companies have explored different solutions to partially address the above-mentioned gap by utilizing external computational power and storage resources provided by either the Cloud or the user's smartphone. Manufacturers have started adding built-in Cloud-based services, e.g., radio, navigation, and software updates, to their state-of-the-art products. Third-party companies have offered different dongles that can be attached to the vehicle and gather various types of data from On-board Diagnostics (OBD) port, which provides a direct access to various sensors and built-in components. Such dongles collect and transmit data (with minimal on-dongle processing) to \textit{smartphone or the Cloud} (either directly or through the smartphone) for further processing. The majority of OBD-connected products support a single or a small set of very basic service(s), such as locking/unlocking doors or closing/opening windows. Recently, a few companies (for example, Mojio \cite{MOJIO}) have introduced new approaches to support multiple applications using a single OBD-connected dongle. Such a dongle transmits raw data to the smartphone/Cloud and enables developers to build on-smartphone or on-Cloud applications, however, it does not offer in-vehicle processing due to resource limitations. 

Despite advantages that on-smartphone or on-Cloud (either manufacturer-enabled or dongle-based) applications offer, their acceptance and application domain \textit{are still very limited due to four fundamental reasons (see Section \ref{SHORT_C} for further discussions):} 

\noindent \textbf{1. Lack of programability:} Typically, vehicle manufacturers do not allow third-party developers to build customized vehicular applications at all or offer limited APIs, e.g., only for entertainment technologies. Vehicles currently have several embedded systems, commonly referred to as Electronic Control Units (ECUs). However, ECUs are designed for and optimized to support basic vehicular operations, such as anti-lock braking system and adaptive cruise control, and are not capable of handling customized applications. 

\noindent \textbf{2. Drawbacks of wireless connectivity:} Vehicle-to-Cloud/smartphone connectivity is not reliable for several (e.g., safety-related) applications due to its potential unavailability and intolerable round-trip delay time. Furthermore, transmitting the huge amount of data needed for data-dominant applications, e.g., traffic sign detection is not cost-efficient.  
    
\noindent\textbf{3. Performance limitations:} Several applications must offer a real-time response, and thus, require in-vehicle processing. Dongles and built-in computing units have limited resources and cannot host computational-heavy applications. Users' smartphones may offer extra resources, however, imposing computational-heavy operations to them significantly increases their power consumption, leading to user inconvenience.
    
\noindent \textbf{4. Public security/privacy concerns:} Add-on dongles do not use strong security mechanisms due to resource constraints, and as a result, they suffer from several security attacks, e.g., the attacker can remotely disable the breaking system \cite{SEC_ATTACK_D}. Furthermore, third-party dongles can transmit a variety of private information to the Cloud, and thus, their introduction has led to public privacy concerns \cite{PRI_ATTACK_D1,PRI_ATTACK_D2,ELASTIC}. For example, Elastic Pathing \cite{ELASTIC}, published in Ubicomp 2014, has shed lights on how insurance companies can infer the user's location by processing the vehicle's speed.

%% CHECK COMPUTATIONAL VS COMPUTATION
%ORDER %>> SmartCore>> Personal >> CLOUD >> ADD-ON

We envision \textit{an interoperable add-on solution} (i.e., solution that imposes minimal design modification to vehicle manufacturers and third-party vehicular companies) as key to enabling a proactive approach to offer new vehicular applications. As discussed later in Section \ref{DESGIN_CONSIDERATION}, an in-vehicle programmable add-on, that imposes no design change to the vehicle, can offer a holistic solution to address the above-mentioned shortcomings of previous smartphone-/Cloud-based services. In this paper, we present a novel reference architecture for vehicular application  development that relies on four fundamental components: (i) SmartCore, \textit{a privacy/security-friendly programmable} OBD-connected dongle that can host multiple applications \textit{in the vehicle}, (ii) the user's personal devices that provide additional resource to applications running on SmartCore and/or enable the user to control them (via a graphical user interface), (iii) Cloud servers, which provide extra resources, keep application installation packages, and offer remote software update, and (iv) add-on modules that enable adding extra input/output devices and computational/storage units to SmartCore if needed. \textit{Based on this architecture, we propose an application development framework for vehicles, called ProCMotive, which enables developers and researchers to rapidly prototype and deploy customized vehicular applications}. 

SmartCore is the core component of ProCMotive and aims to partially push computational/storage resources from the Cloud to the vehicle. In particular, it: 
\begin{itemize}

    \item attaches to the OBD port and replicates a similar interface for third-party dongles. This ensures interoperability, i.e., after adding SmartCore, both the vehicle and previously-designed OBD dongles can resume their regular functionalities. 
    
    \item potentially enables the development of various novel (in particular, latency-sensitive and data-dominant) vehicular third-party applications. It exploits its in-vehicle computational/storage resources to either fully host lightweight latency-sensitive applications or partially implement data-dominant and computationally-heavy applications.  
    
    \item acts as a gateway for third-party OBD-connected dongles. It enables real-time monitoring of other OBD-connected dongles to detect and address any malicious activities (e.g., launching a denial-of-service attack or stealing private information) initiated by them. 
    
    \item offers a context-aware access control scheme that enables the user to decide what information he wants to share with each in-vehicle application or OBD-connected dongle with respect to the current context. 
       
    \item implements a set of privacy-friendly data manipulation functions that aim to minimize the amount of private information leakage by removing inessential parts of data before sharing them with third-party applications and untrusted OBD-connected dongles. 
\end{itemize}

Our key contributions can be summarized as follows:
\begin{enumerate}
   \item We discuss fundamental shortcomings of existing OBD-based add-ons and briefly describe how the proposed approach intends to address them. Furthermore, we suggest a list of additional goals for ProCMotive and justify why each goal is desired. 
   \item We present a reference architecture that comprehensively specifies the functionalities and communication capabilities of its fundamental components (SmartCore, the user's personal devices, Cloud servers, and add-on modules). This architecture has been proposed to address shortcomings of previous OBD-based approaches, while taking suggested design goals into account.
   \item Based on the reference architecture, we design and implement an application development framework that enables developers/researchers to build new vehicular applications.
   \item Using the prototype implementation of ProCMotive, we design and implement two vehicular applications, namely, Amber Response and Insurance Monitor, which are \textit{either completely or partially} hosted on SmartCore.  
   \item We evaluate the applications using real-world data and comprehensively compare them with the state-of-the-art technologies.
\end{enumerate}

The rest of this paper is organized as follows. Section \ref{DESGIN_CONSIDERATION} describes shortcomings of previous smartphone-/Cloud-based approaches, discusses how ProCMotive addresses them, and provides the additional design goals. Section \ref{REF} presents the reference architecture. Section \ref{PROT} explains how we have designed and developed a prototype of ProCMotive based on the reference architecture. Section \ref{APPS} describes two novel applications that we have proposed and implemented based on ProCMotive and evaluates them. Section \ref{RELATED} discusses the related work. Eventually, Section \ref{CONC} concludes the paper. 

\section{Design considerations}
\label{DESGIN_CONSIDERATION}
In this section, we first discuss common shortcomings of previous smartphone-/Cloud-based approaches in detail and briefly discuss how ProCMotive aims to address them. Second, we describe additional design goals, that we considered while designing ProCMotive, and the rationale behind each of them.  

\subsection{Addressing shortcomings of previous approaches}
\label{SHORT_C}
As briefly mentioned in Section \ref{INTRO}, previous approaches have several shortcomings that limit their scope of applications and acceptance. Next, we describe these limitations in more detail and discuss how ProCMotive addresses them. 

\subsubsection{Lack of programability}
State-of-the-art vehicles utilize a compound of ECUs and on-board buses. They incorporate several (up to 100) ECUs, which host vehicle-specific software. ECUs provide in-vehicle resources to enables a variety of basic vehicular operations, such as anti-lock braking system and adaptive cruise control \cite{ECU_0}. Vehicle manufactures have supported ECUs programming and tuning to enhance the vehicle performance even after its initial sale or fix software bugs if needed \cite{ECU_PROGRAM}. However, these built-in computational resources do not offer the flexibility provided by general-purpose computing units: \textit{they cannot be easily reprogrammed to host third-party vehicular applications}. This limitation has been imposed by manufacturers to ensure the quality of service (QoS) and reliability of critical (in particular, safety-related) operations handled by ECUs. Therefore, despite the existence of in-vehicle computational resources, utilizing them to implement customized vehicular applications is neither simple nor recommended. Some manufacturers have started offering APIs to application developers, however, these APIs are very limited and only target a small application domain, in particular entertainment applications. \\
\noindent\textit{\textbf{How does ProCMotive enable programability?}} In the proposed architecture, SmartCore brings extra computational resources to the vehicle, offering a platform that can be used to host a variety of vehicular applications. Since SmartCore is connected to the OBD port, it can access several types of sensory data collected by the vehicle's built-in sensors and communicate with ECUs if necessary. 

\subsubsection{Drawbacks of wireless connectivity}
Here, we discuss the issues associated with the use of vehicle-to-Cloud and vehicle-to-smartphone wireless connectivity.\\
\noindent \textbf{1. Unavailability of wireless connectivity:} Using cellular connectivity to transmit the data from the vehicle to the Cloud will result in the unavailability of the Cloud-based services when the cellular connectivity is not available, for example, when the vehicle goes through a tunnel. Furthermore, if a dongle uses the smartphone to transmit the data to the Cloud, both vehicle-to-smartphone and vehicle-to-Cloud communications become unavailable when the smartphone dies. These will result in the interruption of vehicular services, and significantly limit their applicability. In particular, safety-related applications that need a \textit{reliable continuous stream} of data, e.g., collision detection and security attack detection, cannot completely rely on wireless connectivity. Thus, such applications should be implemented (at least partially) on the vehicle itself. \\
\noindent \textbf{2. Intolerable response time:}
Several vehicular applications need real-time decision making, for example, traffic sign detection and collision prediction, and therefore, may not tolerate the response time offered by the Cloud (i.e., time needed for transmitting the data from the vehicle to the Cloud, processing them on the Cloud, and getting the response back from the Cloud). Such applications should be implemented using close-to-the-vehicle computational/storage resources.

\noindent \textbf{3. Limited cellular data and bandwidth:} Data-dominant applications (in particular, image processing-based collision detection or a sign detection algorithm) capture and process a huge amount of data (up to tens of GBs of data every day). For such applications, transmitting the raw data to the Cloud is not cost-efficient as demonstrated later in Section \ref{APP_1_EVAL}. Moreover, if each data chunk is huge (e.g., a high-resolution image collected from an on-vehicles camera), sending it to the Cloud or the user's smartphone may be time consuming, leading to an intolerable end-to-end application response time. 

\noindent\textit{\textbf{How does ProCMotive address drawbacks of wireless connectivity?}} 
Additional resources offered by SmartCore enables developers to implement applications (either partially or completely) on SmartCore, minimizing the need of using wireless connectivity for data transmission. ProCMotive enables application developers to implement their applications along vehicle-to-Cloud continuum by simultaneously utilizing resources available on SmartCore, nearby personal devices, and the Cloud. Close-to-the-user devices can process a huge amount of data without accessing Cloud resources and only transmit the data over the Internet if necessary.

%As demonstrated later in Section \ref{APPS}, to take the full advantage of both in-vehicle (on SmartCore) and external resources (offered smartphone or the Cloud), the developers can move critical operations to the vehicle and still maintain less critical services on the Cloud. 

\subsubsection{Performance limitations}
Several applications must offer real-time responses, and thus, require in-vehicle processing. Already-in-market dongles and built-in computing units have limited resources and cannot host computational-heavy applications. Users' smartphones may offer extra resources, however, imposing computational-heavy operations to them significantly increases their power consumption, leading to user inconvenience.

\noindent\textit{\textbf{How does ProCMotive resolve performance limitations?}} 
SmartCore brings additional computational/storage resources to the vehicle. Moreover, it minimizes data transmission overheads associated with the use of smartphone-/Cloud-based services since it enables local computation on the data. These enable the implementation of a variety of low-latency/real-time applications on SmartCore. Indeed, SmartCore can run such applications with partially (or even without) utilizing either the user's smartphone or the Cloud. 

\subsubsection{Public security/privacy concerns}
Here, we discuss security and privacy concerns of previously-proposed approaches.\\ 
\noindent \textbf{1. Security threats:} Vehicles are interesting targets for attackers due to their mission-critical operations. Any security attack against vehicles, especially large-scale attacks, may lead to life-threatening consequences. As further discussed later in Section \ref{APP_2}, \textit{the federally-mandated OBD port offers an unprotected standard interface} that can be exploited by attackers to take the control of mission-critical components, e.g., braking system. It has been shown that attackers can launch a multitude of well-known security attacks against the vehicles using OBD port, ranging from Denial of Service (DoS) attacks to packet sniffing \cite{SECURITY_EXP}. Several already-in-market OBD dongles are vulnerable to well-known security attacks, offering a valuable opportunity for attackers to remotely take the control of several components embedded in the vehicle \cite{PRI_ATTACK_D1}.\\
\noindent \textbf{2. Private information leakage:} Since OBD interface offers a full access to all OBD-connected dongles, they can potentially collect a variety of sensitive information (e.g., sensory readings, GPS coordinates, and the vehicle's identification number and model) without the user's permission, leading to several privacy concerns \cite{ELASTIC,PINME}. It has been shown that an insurance company can potentially track all user movements (and also extract several locations of interest) using the vehicle's speed collected from OBD port \cite{ELASTIC}.

\noindent\textit{\textbf{How does ProCMotive enhance security/privacy?}} SmartCore offers sufficient in-vehicle resources to support strong cryptography mechanism (for example, Advanced Encryption Standard encryption \cite{AESBOOK}) needed for protecting wireless communications to/from the vehicle, limiting remote wireless attacks. Moreover, it acts as a gateway that monitors the behavior (e.g., the rate and type of requests) of other OBD-connected dongles to detect and block malicious activities initiated from them. To address privacy concerns, ProCMotive enables two solutions. It offers context-aware access control that enables the user to decide when, where, to what extent, and under what conditions, he wants to share the data collected from OBD with other OBD-connected dongles. Moreover, it implements a set of functions (referred to as privacy-enhancing functions) that manipulate the raw data before sharing it with third-party dongles or applications. 

\subsection{Additional design goals}
\subsubsection{Interoperability}
As mentioned in Section \ref{INTRO}, SmartCore acts as a gateway for other third-party OBD-connected dongles. It connects to the OBD port and replicates a similar interface that can be used by other third-party OBD-based devices, e.g., insurance dongles. Such devices can send their requests to the OBD port \textit{only through SmartCore}, while SmartCore is actively enforcing appropriate security and privacy policies. This ensures interoperability: if an OBD-based dongle complies with the policies, it can perform its regular operations without any design modification, despite the attachment of SmartCore to the OBD. Interoperability is essential to minimize the additional costs associated with the use of ProCMotive and maximize its acceptance. 

\subsubsection{Extensibility} 
It is desired to implement ProCMotive so that its application domain can be extended in future with minimal design modifications. SmartCore offers short-range wireless connectivity (Bluetooth and WiFi), along with several Universal Serial Bus (USB) ports, so that additional input/output, storage, and computing devices can be easily added to the architecture if needed. For example, developers can process the sensory data gathered from the vehicle, along with data collected using add-on input devices (e.g., a camera), to design and develop novel vehicular applications.

\subsubsection{Virtualization}
Virtualization, i.e., creating multiple isolated containers to host different applications on the same operating system (OS), is highly desired to ensure the security. A containers is an abstraction at the application layer that maintains application packages (i.e., codes and their dependencies). SmartCore hosts several third-party applications at the same time, and it uses a separate container for each application. This ensures that an application can neither see nor affect applications running in other containers. Moreover, each container has its own network stack, and therefore, it does not have privileged access to the sockets or interfaces of another container \cite{DOCKER}. 
%For our prototype, we have implemented a middleware that enables virtualization managements, along with several other functionalities (for example, context-aware access control, OBD management, and APIs for accessing sensory data), as described in Section \ref{PROT}. 

%\subsubsection{Collaborative reasoning}

\subsubsection{Remote update} To enhance the performance and security and provide regular fixes for features that are not working as intended, it is essential to offer remote software update. To ensure user convenience, an over-the-Internet remote update is highly desired. In ProCMotive, Cloud servers will host software updates (e.g., the latest version of applications, middleware, and OS) and regularly inform the user if a new update is available. 

%As described in Section \ref{}, in our prototype implementation, the user receives a notification message on his smartphone as soon as an update is available for either the applications installed on the SmartCore or the middleware that manages the applications.    

\section{The reference architecture} 
\label{REF}
\begin{figure*}[h]
\centering
\includegraphics[trim = 20mm 40mm 55mm 17mm ,clip, width=450pt,height=225pt]{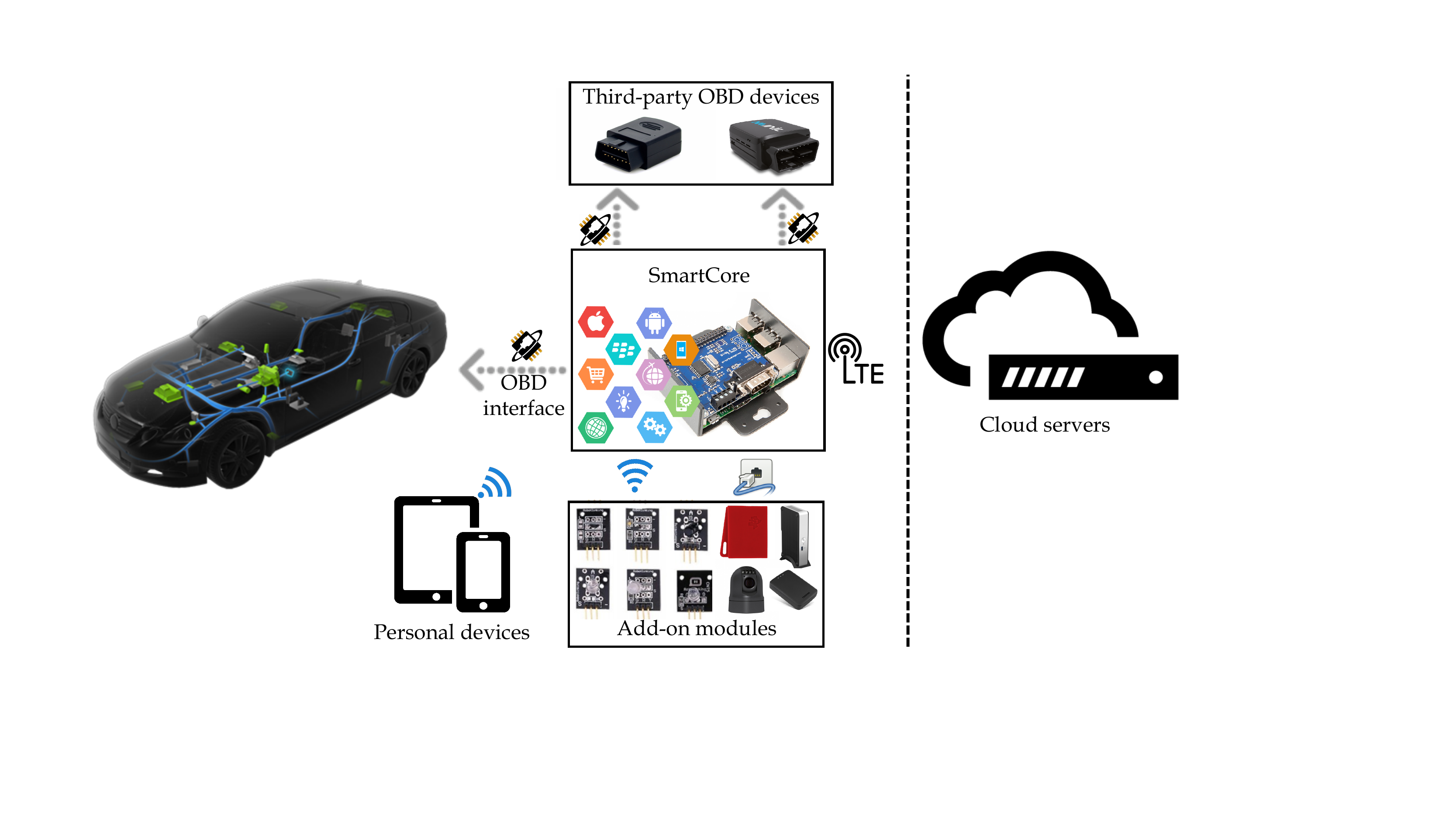}
\caption{An architectural overview of ProCMotive: it consists of SmartCore, personal devices, Cloud servers, and add-on modules. To ensure security, in this architecture, all communication channels (expect OBD-based links) should be encrypted.}
\label{fig:architecure}
\end{figure*}

In this section, we present an architectural overview of ProCMotive. The proposed architecture is motivated by the insight that close-to-the-user computation can open up new opportunities \textit{for addressing various security/privacy concerns associated with the use of Internet-connected vehicles, enhancing the performance of vehicular applications, and enabling new applications that were not feasible before using previous architectures}. Fig. \ref{fig:architecure} presents the proposed reference architecture. As illustrated in this figure, ProCMotive consists of four main components, namely SmartCore, personal devices, Cloud servers, and add-on modules. These components can communicate with each other via various communication channels. To ensure security, in this architecture, all communication channels (expect OBD-based channels that are implemented based on a federally-mandated guideline) can be encrypted. Next, we describe different components of the reference architecture.

%%%\subsubsection{Wireless connectivity} >> TALK BREIFLY HERE

\subsection{SmartCore} 
SmartCore is an OBD-connected dongle that brings sufficient computational power and storage capacity to the vehicle to support several fundamental operations. In the proposed architecture, SmartCore is connected to the OBD port for two main reasons. First, it can access various types of sensory (e.g., coolant temperature, engine RPM, ambient temperature), and non-sensory data (e.g., GPS coordinates, the vehicle's make and model) from the vehicle. Second, it can be powered through this port by accessing the vehicle's battery. Next, we list and briefly describe the fundamental operations of SmartCore.

\subsubsection{Data collection}
SmartCore can collect data needed for various vehicular applications from two main sources: the sensors embedded in the vehicle and add-on sensors. OBD interface enables the SmartCore to access various built-in components including sensors. SmartCore can request different sensory data by sending their corresponding diagnostics parameter IDs (PIDs), which are supported by vehicle manufacturers to facilitate diagnostics. Moreover, add-on input sensors can be connected to SmartCore over WiFi or Bluetooth, providing additional information about the environment.

\subsubsection{On-vehicle data processing}
SmartCore has sufficient resources to perform a wide range of data processing (in particular, privacy-enhancing, data compression, and data analytics) algorithms in the vehicle. Depending on the available resources, performance requirements, and QoS guarantees, applications can be partially/fully implemented on SmartCore. In-vehicle processing opens up a new opportunity for developing several new applications. For example, consider a sign detection algorithm that aims to recognize the traffic signs by processing the images captured from the environment. If the vehicle manufacturer does not support this application by default, incorporating it into already-in-market vehicles is not feasible due to the shortcomings of previously-proposed architectures as described in Section \ref{DESGIN_CONSIDERATION}. However, SmartCore enables in-vehicle image processing for such an application, minimizing the vehicle-to-Cloud transmission overhead and offering short response time. Moreover, as demonstrated later in Section \ref{APPS}, it can be used to implement privacy-enhancing algorithms that remove inessential portions of raw data (e.g., the whole image) before transmitting it to the Cloud or sharing it with other OBD-connected devices, e.g., insurance dongles.  

\subsubsection{Access control}
While various access control schemes have been proposed for personal devices (smartphones and tablets) \cite{ConXsense,SMA1,SMA2}, they have been neither well-established nor well-studied in the domain of Internet-connected vehicles and vehicular applications. The OBD protocol itself does not offer any access control solution to specify when, where, and to what extent the sensitive data can be gathered from the OBD. In order to prevent forming a monopoly in the auto repair business, vehicles manufacturers are mandated by law to provide full access to built-in components via OBD port. Although OBD-connected dongles can access various sensors and components embedded in the vehicle to enable new vehicular applications, their usage can lead to serious security and privacy concerns if their access level is not limited. SmartCore offers an access control scheme to limit the access level of (i) applications hosted on the SmartCore and (ii) third-party OBD-based dongles. It continuously monitors the behaviors of hosted applications and third-party dongles and ensures that they comply with a set of access control policies. SmartCore supports two types of policies: predefined policies and context-aware user-defined policies, as described next.\\
\noindent\textbf{Predefined policies:} Upon the installation of an application or the attachment of a new dongle to SmartCore, a set of predefined policies are assigned to the application/dongle. These set of policies are determined based on two parameters: the vehicle's specifications (e.g., vehicles' manufacturer, make, and model) and the specification of the application/dongle (e.g., the application's intentions or the manufacturer/model of the dongle). The vehicle's specification can be directly obtained from the OBD port. It is used to take the vehicle's manufacturer-reported OBD issues and specific characteristics into account. For example, the attachment of any OBD dongle to a Ferrari 430 will disable its Traction Control System \cite{FERRARI}. Thus, for this vehicle, the default access level of applications/dongles should ensure that the OBD port cannot be accessed when the car is moving. The specification of the application/dongle is used to determine its expected access level based on its intended operations. For example, it is sufficient for an insurance dongle to only access a subset of the vehicle's sensors, e.g., the speedometer and odometer. \\
\noindent\textbf{Context-aware user-defined policies:} 
Although predefined policies provide basic protection against different security/privacy attacks, it is unlikely that the their privacy and security implications would be fully understood by regular users. Indeed, users commonly under-/over-estimate the level of protection that these policies provide \cite{ConXsense}. To take users' preferences into account, we included a domain-specific context-aware access control scheme in SmartCore. Context-aware user-defined policies offer the potential to correctly reflect the user's security/privacy preferences. However, if it is not user-friendly, the amount of essential user effort needed to initialize, modify, and maintain a comprehensive set of context-dependent policies is high \cite{ConXsense}. A rich set of contexts enables the user to define fine-grained policies, however, it is well-known that regular users are not willing to spend significant amounts of time to adjust the policies with their preferences. In addition, it is questionable, whether users are capable of understanding the implications of their policy settings \cite{ConXsense}. Thus, it is desired that SmartCore offers a user-friendly policy managements system, while enabling users to define/modify various policies with respect to a rich set of domain-specific high-level contexts (e.g., whether the vehicle is involved in an accident).
\subsubsection{OBD port management}
In order to enable real-time monitoring of other OBD-based third-party dongles, such dongles are only allowed to indirectly access OBD port through the interface implemented by SmartCore. SmartCore should isolate third-party OBD-based dongles (i.e., they can neither directly communicate with the vehicle's OBD port nor see each other), monitor and process commands/message initiated from them, and responds to such commands/message (with respect to certain security/privacy policies specified by the access control scheme). Port management enables this isolation and handles requests sent to or received from the vehicle's OBD port.

\subsection{Personal devices}
Modern personal devices (e.g., smartphones and tablets) have become a vital part of our everyday lives. They are equipped with many compact built-in sensors (e.g., accelerometers, magnetometers, and barometers), various communication capabilities (e.g., WiFi, LTE, and Bluetooth), powerful microprocessors, and high-volume storage in order to support a variety of applications \cite{PINME}. In addition to enabling such applications, their spare resources \textit{can be potentially utilized to offer additional resources (e.g., computational power) and inputs/outputs (e.g., sensors) to applications running on SmartCore}. For example, if a developer wants to build an automatic headlight control application (i.e., an application that can automatically turn on the vehicle's headlights based on the existing ambient light) on a vehicle that does not support this functionality. He can potentially utilize the ambient light sensor embedded in the user's smartphone to sense the ambient light and then launch appropriate controlling commands to control the vehicle's headlight using SmartCore. Furthermore, its display can offer a user-friendly interface, which can be used to control various functionalities of SmartCore, as further described later in Section \ref{ANDROID}.    

\subsection{Cloud servers}
In the proposed architecture, Cloud servers are envisioned to have three fundamental responsibilities: (i) maintaining application packages, (ii) offering additional computational/storage resources that can be used to partially/fully implement an application on the Cloud, and (iii) enabling remote update of the software, in particular, applications installed on SmartCore, and the underlying middleware and/or OS. 

\subsection{Add-on modules}
These modules are either additional input/output devices (e.g., camera and sensors) or computing/storage units that can be connected to SmartCore via WiFi, Bluetooth, or wired connectivity. For example, a vehicle-mounted camera can gather valuable information about the vehicle's surroundings, enabling a variety of image processing-based applications. In Section \ref{APPS}, we discuss an instance of novel applications that can be enabled using an add-on module, a vehicle-mounted camera. \textit{In the proposed architecture, add-on modules offer extensibility}.

\section{Prototype implementation}
\label{PROT}
Based on the proposed reference architecture and intended operations of its components, we designed and developed a prototype for ProCMotive. The prototype implementation offers a framework that provides the backbone for vehicular application development, considering various domain-specific challenges: programability, wireless connectivity, and performance shortcomings, along with security/privacy concerns. Next, we discuss the implemented software components and underlying hardware/infrastructure.  

\subsection{Software components}
The prototype implementation consists of three main software components: (i) CARWare a middleware (i.e., a software that acts as a bridge between the native OS and applications) that enables SmartCore's intended operations, (ii) an Android application that enables the user to control different vehicular applications and manage access control policies using his smartphone, and (iii) a trusted web server on the Cloud that enables the user to download/update vehicular application packages. Next, we discuss these components in detail.  

\subsubsection{CARWare: The core middleware}
\begin{figure}[h]
\centering
\includegraphics[trim = 80mm 65mm 95mm 45mm ,clip, width=250pt,height=130pt]{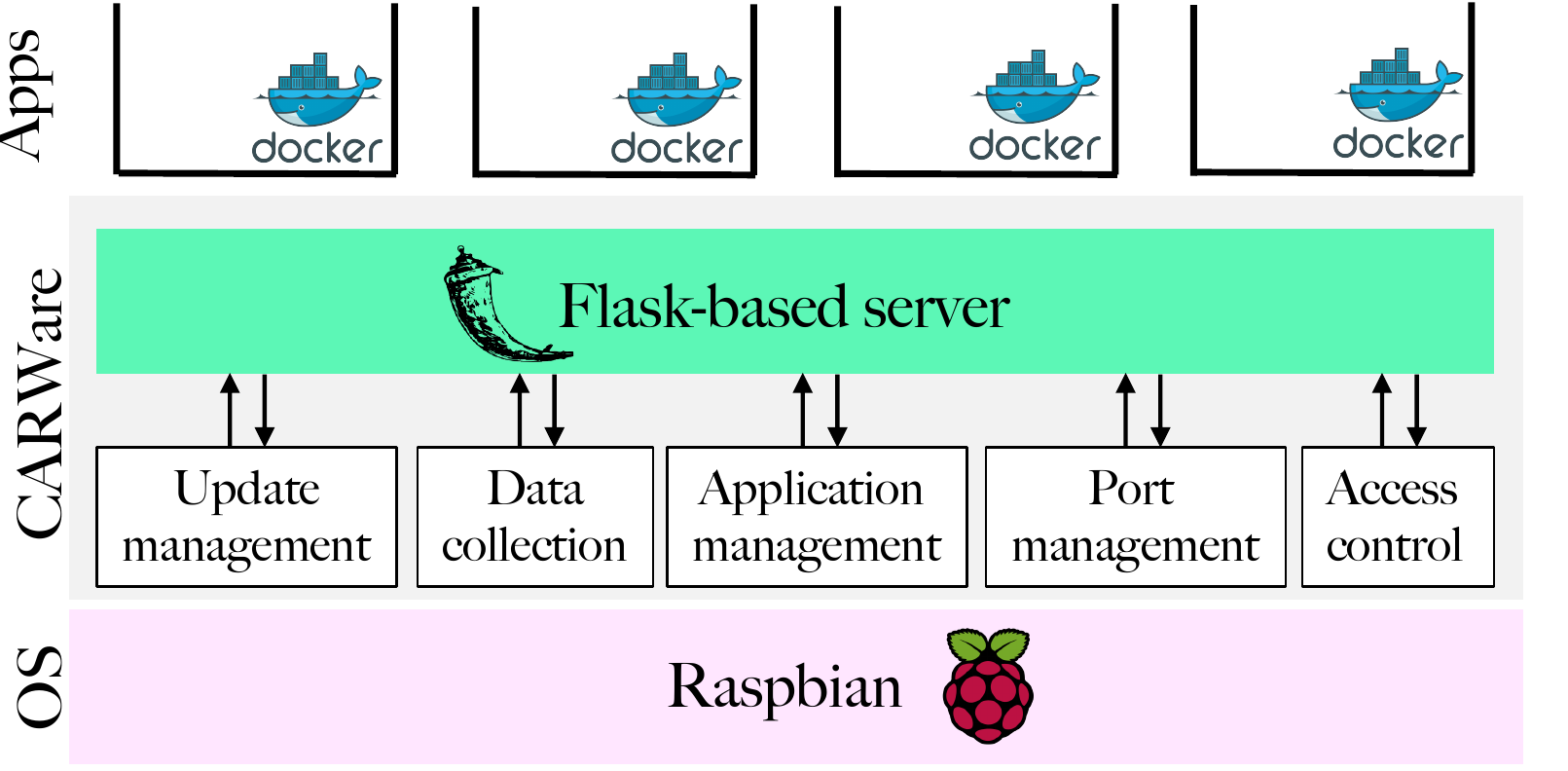}
\caption{CARWare: The middleware comes between the OS and the application layer and provides various APIs for update management, data collection, application management, port management, and access control, which can be accessed through its Flask-based web server.} 
\label{fig:CARWare}
\end{figure}

As the core of reference architecture, SmartCore offers several functionalities. We have designed and implemented CARWare to enable the intended functionalities of SmartCore discussed earlier in Section \ref{REF}. CARWare comes between the native OS (Raspbian \cite{Raspbian} in our prototype) and the application layer (Fig. \ref{fig:CARWare}). It enables remote update, data collection (from both OBD port and add-on inputs), application management, OBD port management, and access control and provides a RESTFul API through which it handles various requests created by applications and returns a response in JSON format (i.e., an open-standard file format that uses human-readable text). Next, we describe the supported functionalities of CARWare and briefly describe the APIs that it provides. 

\noindent\textbf{Update management:}
The update management provides various functions needed for remotely updating application packages and the CARWare. In our prototype, we have used the full functionally of API to develop a \textit{trusted Android application (with administrative privileges)} that can download, install, and run the last version of application packages and CARWare stored on the Cloud if an update is necessary. Although update management enables updating both applications and CARWare, for \textit{third-party applications}, its API is very limited: it only enables such applications to update themselves. For security reasons, we do not allow a third-party application to neither update other applications nor CARWare. \\
\noindent\textbf{Data collection:}
CARWare offers an API that can be used to facilitate data collection from the vehicle's sensors and SmartCore's add-on inputs. The current implementation of the API enables accessing over 30 types of the sensory data from the vehicle, along with different types of data provided by three add-on inputs: a vehicle-mounted camera (Raspberry Camera Module V2 \cite{RASP_CAM}), a set of sensors (TI Sensor Tag \cite{TI_TAG} that has built-in accelerometer, magnetometer, temperature, and air pressure sensors and is connected to SmartCore via Bluetooth Low Energy), and a GPS receiver (USGlobalSat GPS receiver \cite{GPS_REC}). In order to fetch sensory data, the vehicular application, that runs in a container, communicates with the web server running within CARWare. For data collection, the application creates a request including its unique credentials (an application identifier and a token) along with the description of data that are needed (for each data type, the request includes two fields: source of data, i.e., from the 'vehicle' or 'add-on' inputs, and type of data, e.g., acceleration or engine RPM). Upon the arrival of a request, CARWare first checks if the request complies with access control policies. If so, it reaches the requested data source, collects the data, and returns a response including the data (in JSON format) to the application. Otherwise, it rejects the request. \\
\noindent\textbf{Application management:} In-vehicle application hosting requires fine-grained application management. Application management enables the user to (use an Android application developed based on its API and) download a vehicular application from Cloud server to the SmartCore, run the application (inside a container separated from other applications), pause all processes involved in the application, and completely halt the application. Using containers allows independent isolated applications to run simultaneously within a single OS, avoiding the overhead of starting and maintaining several virtual machines. In our prototype, we utilize Docker technology \cite{DOCKER} to create our isolated containers. Docker is one of the world's leading software container platform that facilities the repetitive tasks of building containers and configuring development environments \cite{WHAT_DOCKER}. 
\\
\noindent\textbf{Port management:}
CARWare offers an API to enable applications to control the OBD port if needed. In particular, it provides four functions: port block, rate adjustment, probing a dongle, and sending a request. Using port management, an application can (i) block all requests coming from an OBD-connected dongle (given its unique identifier), (ii) set the maximum expected rate of OBD requests initiated from an application/dongle, (iii) capture/monitor all the packets initiated from a dongle, and (iv) create and transmit an arbitrary OBD request. As demonstrated later in Section \ref{APPS}, using this API, developers can easily design security/privacy protection applications, which can take the control of OBD port upon detection of a malicious activity.

% CHECK CONSISTENCY FOR OBD-based third-party devices 
%% ACCESS CONTROL REMOVES DOS AS WELL: 00 PID cannot be sent by anyone. % CHECK CONSISTENCY FOR OBD-based third-party devices 
%% ACCESS CONTROL REMOVES DOS AS WELL: 00 PID cannot be sent by anyone. 
%RATE
% RELAY 

\begin{figure}[h]
\centering
\includegraphics[trim = 140mm 53mm 95mm 60mm ,clip, width=250pt,height=165pt]{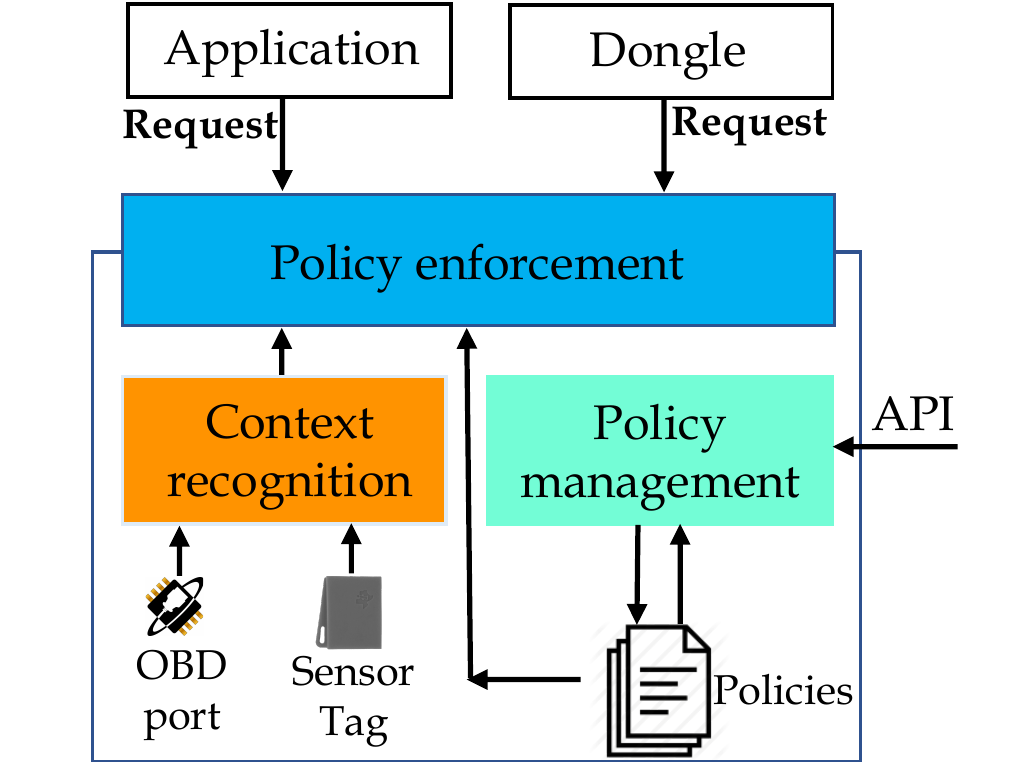}
\caption{The access control component consists of three subsystems: policy management, policy enforcement, and context recognition} 
\label{fig:Access}
\end{figure}

\noindent\textbf{Access control:}
The access control component consists of three subsystems: policy management, policy enforcement, and context recognition that closely collaborate with each other (Fig. \ref{fig:Access}):\\

\textit{1. Policy management:} This subsystem is responsible for getting the feedback from the user and enabling the user to enforce his security/privacy preferences. It provides an API which can be used to add, edit, and remove context-aware user-defined policies for each application or third-party dongle attached to SmartCore.\\ 
\textit{2. Policy enforcement:} It ensures that all applications and third-party OBD dongles always comply with both predefined and user-defined policies. For each request generated from an application or dongle, if the request is authorized, it lets the request to proceed; otherwise, it blocks the request, i.e., the request is neither processed by SmartCore nor forwarded to the OBD port. \\
\textit{3. Context recognition:} Context recognition supports a list of contextual information (it continuously detects the current contexts), enabling the user to set his preferences with respect to this information. Next, we describe different type of contextual information supported in the prototype. \\
\noindent \textbf{Operational contexts:} In the prototype, the context recognition supports two contextual information related to the operation of the vehicle: \textit{vehicle status and health status}. It continuously detects whether the vehicle is idle or moving. This allows the user to limit applications/dongles based on the current status of the vehicle. For example, the user can set a policy to disable all diagnostic OBD dongles (that may send safety-critical commands to the vehicle) when the vehicle is moving. This can potentially prevent several life-threatening security attacks (e.g., disabling the braking system \cite{SEC_ATTACK_D}) even if the dongle is hacked and can be controlled by a remote attacker. Moreover, it detects the engine's health status (e.g., checks whether the check engine light is on/off). Several insurance companies, e.g., MetroMile \cite{METRO}, offer dongles that are also able to read all in-vehicle data, find fault codes, and describe how the user can address the issue. By setting policies based on the health status, the user can allow such dongles to only access diagnostic data upon the appearance of a fault. 

\noindent \textbf{Situational contexts:}
In the prototype, two types of situational contexts are defined: \textit{involvement in an emergency and the presence of an external alert message}. Context recognition frequently collects data from sensors embedded in TI Sensor Tag \cite{TI_TAG}, e.g., accelerometers, and on-vehicle sensors, e.g., ABS and airbag sensors, to detect the occurrence of a collision. Moreover, it listens to a trusted communication channel through which trusted alert messages (e.g., from law enforcement) are transmitted to the vehicle. This enables the user to set situational policies. For example, a user may be willing to give an insurance dongle the permission to transmit the location of an accident to the company. Similarly, he may want to allow emergency responders to access his location information when it is asked by a law enforcement agency (see Section \ref{APPS} for an application, called Amber Alert, that needs this type of permission).

\noindent \textbf{Location-based contexts:} There are different scenarios in which the user might like to control the access level of the application/dongle with respect to the current location of the vehicle. For example, the user may be willing to share some information with applications/dongles only when he is in trusted locations. Furthermore, he might want to stop sharing his sensitive information (e.g., GPS coordinates) when he is in specific locations (e.g., his home or office). The current prototype implementation lets the user to set policies with respect to a set of locations of interest: user-defined locations (home and office addresses) and manufacturer-trusted addresses (locations of trusted auto repair shops).

%\noindent \textbf{Time:}
%For the majority of scenarios, e.g., annual vehicle inspection procedure or occasional auto services, a temporary access to OBD port may be sufficient. However, for some scenarios, e.g., using pay-per-mile insurance devices, enabling remote functionalities (for example, turning on/off the vehicle), and locating the car, a long-term access to OBD port is essential. The system can assign a temporary or permanent access level to the device.  All permissions given to a device with temporary access will be automatically revoked after time T\_Access, whereas other permissions will remain in place until the user manually revokes them.  

\subsubsection{Cloud server}
In the prototype, ProCMotive has a trusted Cloud-based web service, called \textit{Vehicular Application Store} that is written in Python based on Flask framework \cite{FLASK} and is hosted on Amazon Web Services (Model t2.2xlarge \cite{AWS}). It offers an API that can be used to: (i) list available vehicular application packages, and (ii) download the last (or a specific) version of an application files (a Docker container \cite{DOCKER} including the application and its dependencies along with a JSON file describing the application and its requirements). Using the API provided by this sever and the API provided by the update management unit of SmartCore, we developed and Android application that allows the user to download, install, and update applications/CARWare, as described later in Section \ref{ANDROID}. 

\noindent \textbf{Note:} As described earlier, the Cloud has a vital role in the implementation of various vehicular applications: it can provide extra computational/storage resources for each application. While designing different vehicular applications based on ProCMotive, we have also used additional resources of the Cloud (see Section \ref{APPS} for more detail, where we describe two vehicular applications developed based on ProCMotive). 

\subsubsection{Android application}
\label{ANDROID}
Using the APIs provided by Vehicular Application Store and SmartCore, we have implemented an Android application that communicates with both the store and SmartCore and offers a user-friendly interface for managing access control policies and vehicular applications running on SmartCore: 

\noindent\textbf{Access control management:} The user can set, modify, delete user-defined access control policies. For each vehicular application or OBD dongle, CARWare maintains an access control file (in JSON format) that specifies its access control policies. The smartphone communicates with the web server within CARWare to reflect user preferences and update the access control files. \\
\noindent\textbf{Application management:} To install and run a new application, the user can select one application from the list of available vehicular applications stored on Vehicular Application Store. When the user intends to download and run an application on SmartCore, his request is sent from the smartphone to the web server within CARWare and is handled as follows: the server communicates with Vehicular Application Store and fetches the application package, it then runs the application in an isolated container. Using the smartphone application, the user can check the status of all vehicular applications running on SmartCore and manage them (pause, halt and remove their containers) if needed.\\

\subsection{The underlying hardware/infrastructure}
SmartCore is implemented based on Raspberry Pi 3 that comes with Raspbian, its native Debian-based computer OS \cite{Raspbian}. SmartCore also utilizes two other hardware components: NETGEAR 4G LTE Modem (Model LB1120) \cite{NETGEAR} with a T-Mobile Prepaid Plan \cite{Tmobile} that enables wireless connectivity over LTE and OBD PiCAN 2 board \cite{PICAN} that provides CAN-Bus capability for the Raspberry Pi. It currently offers three wireless channels: LTE, Bluetooth Low Energy, and WiFi. Raspberry Pi has built-in cryptographic modules that support strong encryption for these communication channels. Thus, to ensure security, all communication channels to/from SmartCore can be encrypted (except the wired OBD-based communication channels).

Furthermore, SmartCore currently supports three add-on modules: (i) TI Sensor Tag \cite{TI_TAG}, a Bluetooth-enabled sensory unit that includes various sensors such as accelerometer, magnetometer, and air pressure, (ii) Raspberry Camera V2 \cite{RASP_CAM}, a vehicle-mounted camera that can capture video frames from the environment, and (iii) USGlobalSat GPS \cite{GPS_REC} that is a GPS receiver. 

We have used a Nexus 5S to test all Android applications developed in this paper and utilized Amazon Web Services (Model t2.2xlarge \cite{AWS}) as the Cloud.

\section{Applications}
\label{APPS}
In this section, we propose two novel applications, which are implemented based on ProCMotive. We evaluate these applications using real-world data and discuss how they benefit from in-vehicle processing (SmartCore).
\subsection{Application 1: Amber Response}
In this section, we discuss a novel application that has been enabled by ProCMotive, which we call Amber Response. 

In the U.S., an Amber Alert is activated when a law enforcement agency has admissible reasons to believe that a child has been abducted and he is in danger of serious life-threatening conditions or death. The Amber Alert system relies on the nearby people to get information about the abduction. It informs the public about the abduction by broadcasting the \textit{make, model, color, and plate number} of the abductor's vehicle to nearby smartphones, enabling the entire community to assist in the safe recovery of the child. It has been shown that this scheme is only slightly effective and may cause user inconvenience (for example, the alert will be sent to all nearby people even if they are not walking/driving and cannot provide useful information). Since the inception of the program in 1996 through 2015, around 43 children, on average, have been safely recovered every year specifically as a result of an AMBER Alert being issued, whereas the average number of abduction in the U.S. is around 800,000 every year (see \cite{AMBER_STAT} for detailed annual statistics). Thus, a more effective alternative system is highly needed. We propose such a system, called Amber Response, and implement it using ProCMotive. Amber Response utilizes a vehicle-mounted camera, that continuously captures several frames per second from the environment, and processes image frames to automatically find the abductor's vehicle (given the database of active Amber Alerts). Different functions of this application can be distributed across SmartCore and the Cloud, as described next. 

\subsubsection{Prototype implementation}
Amber Response application maintains a database of active Amber Alerts, including make, model, color, and plate number of abductors' vehicles.  This database is located on the Cloud server and can be updated by responsible agencies. The application searches through the video frames to find a vehicle whose features match the ones of a record in the database. Upon the detection of a suspicious vehicle, the application sends the vehicle's GPS coordinates to the Cloud server, informing law enforcement agencies. Using ProCMotive, we have implemented three different versions of Amber Response: (i) a Cloud-based, (ii) a SmartCore-based, and (iii) a hybrid version that exploits both Cloud and in-vehicle computation/storage resources. We next describe how we have implemented these three versions. 

\noindent\textbf{Cloud-based:} In this version, SmartCore only collects the data (video frames) and uploads them to the Cloud without modification. After uploading the frames, an on-Cloud server receives and processes them to find a plate number that matches the plate number of a suspicious vehicle in the database. We have utilized OpenALPR library \cite{ALPR} to implement plate detection algorithm on the Cloud. Plate detection algorithm has eight main steps, which are briefly described in Table \ref{table:OPENAPLR}. Implementing the Cloud-based version of Amber Response was potentially feasible using previously-proposed Cloud-based architectures that rely on minimal computational power in the vehicle, and thus, the Cloud-based version can be used as a baseline to compare ProCMotive-enabled (SmartCore-based and Hybrid) implementations with previously-presented Cloud-based proposals for connected vehicles (e.g., Azure-based connected vehicles \cite{AZURE}). As described later in Section \ref{APP_1_EVAL}, despite demonstrating a promising performance, the Cloud-based implementation cannot be utilized in real-world scenarios due to the cost overhead associated with transmitting the data needed for this implementation.

\begin{table}[ht] 
\caption{Different steps of plate detection algorithm \cite{ALPR}} % title of Table 
\centering % used for centering table 
\begin{tabular}{l l} % centered columns (4 columns)
Step & Description\\
\hline\hline
1. Plate detection & Finds potential license plate regions\\
2. Binarization & Converts the plate image into black and white\\
3. Char Analysis & Finds character-sized ``blobs'' in the plate region\\
4. Plate Edges  & Finds the edges/shape of the plate\\
5. Deskew & Transforms the perspective to a straight-on view\\
6. Segmentation & Isolates and cleans up the characters\\
7. Char Recognition & Analyzes each character image \\
8. Post Process & Creates a top N list of plate possibilities\\
\hline %inserts single line 
\end{tabular} 
\label{table:OPENAPLR}
\end{table}

\noindent\textbf{SmartCore-based:} In this version, the application installed on SmartCore frequently (e.g., every 30 seconds) fetches the database of active Amber Alerts to ensure that it maintains the last updated version of the database. It then captures video frames from the camera and runs the plate detection algorithm described above. After extracting all plate numbers from the frames, it searches through the database to find a match and sends a report, including the vehicle's GPS coordinates, to the Cloud if a match is found.

\noindent\textbf{Hybrid:} The hybrid implementation exploits both in-vehicle and on-Cloud resources. In this scenario, Amber Response application has been partially implemented on SmartCore. On SmartCore, it first captures the frames from the camera. Then, it performs a lightweight image processing function to extract all plate areas in each frame (Step 1 in Table \ref{table:OPENAPLR}). Afterwards, for each vehicle in the frame, it estimates the vehicle's color from a small area above its plate (whose size is $\%10$ of the detected plate's area). If the vehicle's color matches the color of one of the suspicious vehicles reported in the database, it transmits the corresponding plate area to the Cloud for further processing. The on-Cloud side of the application, receives and processes the images that only contain the area. Upon detection of a suspicious vehicle, it sends a request to the application installed on SmartCore, asks for current location of the vehicle, and informs the law enforcement agency. 

\subsubsection{Evaluation}
\label{APP_1_EVAL}
Next, we first describe the dataset used to evaluate Amber Response, and then examine and compare different implementations of Amber Response from three perspectives: (i) performance, (ii) cellular data usage, and (iii) privacy leakage.

\noindent \textbf {Dataset:} We downloaded 12 videos uploaded on YouTube that were captured using a camera mounted behind the mirror of a moving vehicle. These videos have the resolution of at least 1080p and frame rate of at least 10 frames per second (FPS). To construct our dataset, we created 72 videos by varying both resolution and frame rate of the downloaded videos. For each original video, the dataset includes six videos with different resolutions, i.e.,  1080p and 720p,  and frame rates, i.e., 1, 5, and 10FPS. Each video is about 10 minutes and the suspicious vehicle, i.e., the vehicle that Amber Response aims to find, appears in a random time in the footage (i.e., for each video, we randomly choose a single vehicle and add its specifications to the database of suspicious vehicles stored on the Cloud). Based on our empirical results, Amber Response can accurately (with the accuracy of $\%100$) detect the abductor's vehicle when the frame rate is equal to or greater than 1FPS. Therefore, in our evaluations, the minimum frame rate is set to 1FPS. 

\noindent \textbf{Comparison:} We first quantitatively evaluate three implementations of Amber Response (Cloud-based, SmartCore-based, and Hybrid) using the above-mentioned dataset. Then, we briefly describe how in-vehicle data processing can enhance the user's privacy in this application. 

\noindent \textit{1. Performance:} In order to compare the performance of the three implementations, we define and report Detection Time Ratio (i.e., $DTR=\frac{T_{detection}}{T_{appearance}}$) that represents how much time each implementation takes for processing one second of the video until it finds the abductor's vehicle. This metric enables us to estimate the delay in the detection of the abductor's vehicles in real-world scenarios: if the suspicious vehicle appears after $T_{appearance}$ seconds (from when the camera starts capturing the video), Amber Response detects it after $T_{appearance}*DTR$. Indeed, it reports the suspicious vehicle with a delay of $T_{appearance}*DTR-T_{appearance}$ seconds to the law enforcement agency.  
For each implementation, the $DTR$ highly depends on frame rate and resolution of the video, and how much processing power is provided by SmartCore. Next, we examine how average DTR changes with respect to these parameters.

\begin{figure}[h]
\centering
\includegraphics[trim = 105mm 60mm 130mm 45mm ,clip, width=230pt,height=185pt]{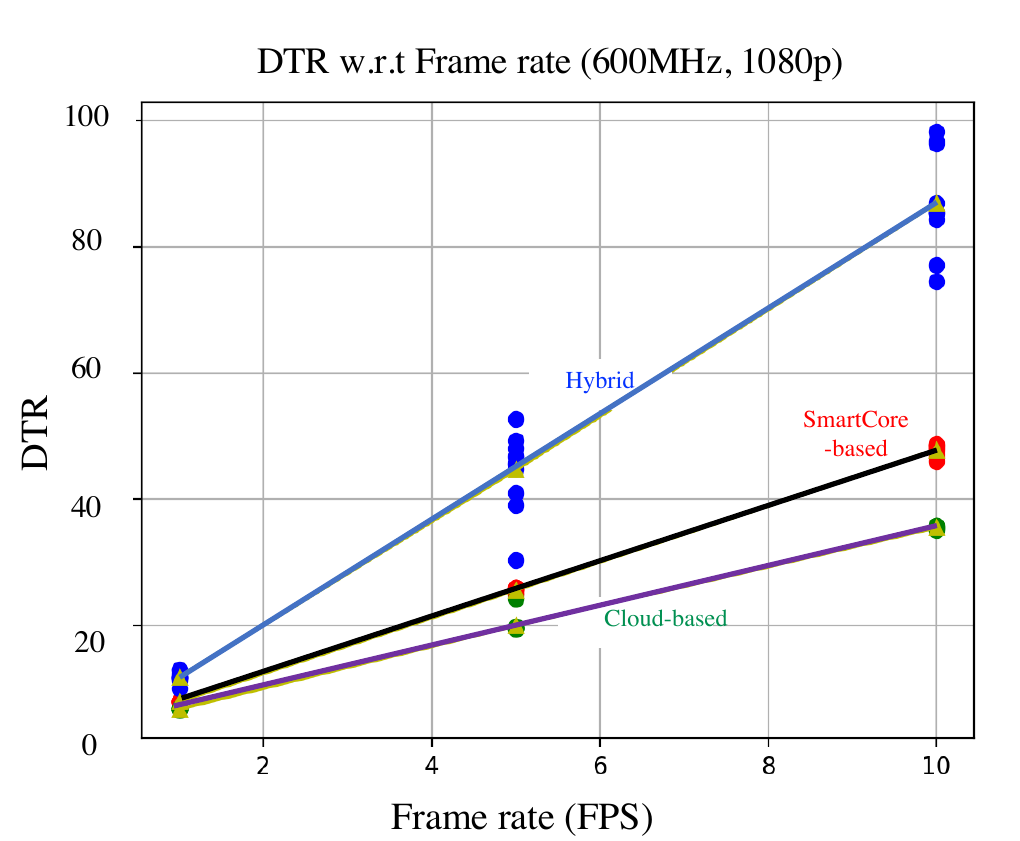}
\caption{DTR with respect to the frame rate: each point represents DTR for a single video from the dataset (lines show how average DTR changes with respect to the frame rate).} 
\label{fig:DTR}
\end{figure}

\noindent\textbf{\textit{Experimental scenario 1:}} In order to evaluate how $DTR$ changes with respect to the frame rate, we examined Amber Response using a subset of the videos (36 videos) in the dataset that have the same resolution (1080p) and we ensured that SmartCore provides similar processing power for all these videos by manually enforcing its CPU to work at $600MhZ$ (on Raspbian \cite{Raspbian}, this can be done by editing ``config.txt'' located at ``/boot/config.txt''). Fig. \ref{fig:DTR} demonstrates how $DTR$ changes with respect to the frame rate for this experimental scenario. As the frame rate increases (and consequently, the image processing algorithm becomes more computationally-heavy), utilizing the Cloud for processing becomes more reasonable from performance perspective, whereas when the frame rate is low (1FPS) all three implementations become similar from performance perspective even though the computational power of SmartCore is much less than that of the Cloud. 

\noindent \textbf{\textit{Experimental scenario 2:}} Furthermore, to evaluate how $DTR$ changes with respect to the resolution of the video, we have repeated a similar experiment: we manually enforced the CPU to work at $600Mhz$ and used the videos with $1FPS$. Table \ref{table:perf720and1080} summarizes the results of this experiment. As expected, for each implementation, a lower resolution video offers a better $DTR$. 

\indent \textbf{A notable observation}: As shown in Fig. \ref{fig:DTR}, in the first experimental scenario, both Cloud-based and SmartCore-based implementations outperformed the hybrid one. For hybrid implementation, $T_{detection}$ highly depends on by how much time (i) on-SmartCore processing, (ii) SmartCore-to-Cloud data transmission, and (iii) on-Cloud processing take. For videos with the resolution of 1080p, the hybrid version of Amber Alert spends a significant amount of time for on-SmartCore processing (Step 1 from Table \ref{table:OPENAPLR} and color detection), and therefore, it is slower than both SmartCore-based one (for which SmartCore-to-Cloud data transmission and on-Cloud processing times are zero) and Cloud-based one (for which on-SmartCore processing time is negligible). \textit{However, when the resolution of input videos is changed to 720p, the hybrid implementation outperformed the SmartCore-based one (Table \ref{table:perf720and1080}).} In this experimental scenario, for the hybrid implementation, on-SmartCore processing takes significantly less time (compared to when the resolution of the input video is 1080p) so that it is reasonable to shift several steps of the plate detection algorithm (Step 2 to Step 8 from Table \ref{table:OPENAPLR}) to the Cloud despite the additional time overhead associated with SmartCore-to-Cloud data transmissions.

\begin{table}[ht] 
\caption{DTR with respect to the resolution (600 MhZ, 1FPS)} % title of Table 
\centering % used for centering table 
\begin{tabular}{l c c c} % centered columns (4 columns)
Resolution & Cloud-based & SmartCore-based & Hybrid \\
\hline\hline %inserts double horizontal lines 
1080p & 6.8 & 8.0 & 11.88 \\
720p & 4.4 & 5.7 & 5.57 \\
\hline %inserts single line 
\end{tabular} 
\label{table:perf720and1080}
\end{table}

\noindent\textbf{\textit{Experimental scenario 3:}} Eventually, in order to examine how the performance of different implementations may vary with changes in computational power, in another experimental scenario, we have manually overclocked SmartCore's CPU to run at $1200MhZ$ and repeated our examination using a subset of videos (with the resolution of 720p and frame rate of 1FPS). Table \ref{table:perf1200} summarizes the results of this experiment. In this experimental scenario, where the computational power of SmartCore is significantly increased, SmartCore-based implementation outperformed both hybrid and Cloud-based implementations, indicating that the additional time needed for SmartCore-to-Cloud data transmissions and on-Cloud processing is more greater than performing all steps (Steps 1-8 in Table \ref{table:OPENAPLR}) on SmartCore. 

\begin{table}[ht] 
\caption{DTR for different implementations (1200 MhZ, 720p, 1FPS)} % title of Table 
\centering % used for centering table 
\begin{tabular}{l c} % centered columns (4 columns)
Implementation & DTR\\
\hline\hline %inserts double horizontal lines 
Cloud-based & 3.6\\
SmartCore-based & 3.0\\
Hybrid & 4.3\\
\hline %inserts single line 
\end{tabular} 
\label{table:perf1200}
\end{table}

\noindent \textit{2. Cellular data usage:} Using our real-world dataset, we have examined how much cellular data, on average, each implementation has used for processing the 10-minute videos in the dataset. For both Cloud-based and hybrid implementations, the data usage highly depends on both FPS (that specifies the data transmission frequency) and resolution of the video (that determines the size of each packet transmitted to the Cloud). Furthermore, for Hybrid implementation, the sparsity of the environment (e.g., how many vehicles are present in each video frame) has an effect on the cellular data usage: in crowded areas, the hybrid implementation of the application transmits more areas of interest to the Cloud. As shown in Table \ref{table:datausage}, the Cloud-based implementation consumes the most cellular data among the three implementations, whereas the SmartCore-based one utilizes the least (i.e., it only transmits the vehicle's GPS coordinates to the Cloud upon the detection of an abductor's vehicle). Hybrid implementation offers $34.8X$ reduction in cellular data usage in comparison to the Cloud one, at the cost of performing a lightweight algorithm on SmartCore. SmartCore-based implementation only occasionally communicates with the Cloud (to receive the updated database of active Amber Alerts and send the location of the vehicle upon the detection of a suspicious vehicle), however, it imposes significant computational overhead to SmartCore by locally processing all images.

\begin{table}[ht] 
\caption{Cellular data usage (1FPS, 720p, 10 minutes)} % title of Table 
\centering % used for centering table 
\begin{tabular}{l c} % centered columns (4 columns)
Implementation  & Cellular data usage (MB)\\
\hline\hline %inserts double horizontal lines 
Cloud-based & 115.0 \\
SmartCore-based & 0.0\\
Hybrid & 3.3\\
\hline %inserts single line 
\end{tabular} 
\label{table:datausage}
\end{table}

\textit{Note that the Cloud-based implementation, that can be also implemented using the previously-proposed Cloud-based architectures described in Section \ref{INTRO}, cannot be used in real-word scenarios due to its high cellular data usage.} For example, if the user only drives for one hour every day, it requires transmitting over 20 GBs of data every month over cellular network (assuming the video is captured at frame rate=1FPS and resolution=720p). This imposes a significant cost overhead to the user (currently, in the U.S., the cost is close to $\$100$ per month for 20GBs).\\
\noindent \textit{3. Privacy leakage:} Transmitting raw images captured from the vehicle to third-party servers, can potentially leak significant private information, including, but not limited to, the specifications of the vehicle (e.g., make, color, and model), the area that the vehicle's owner is travelling through, and the owner's locations of interest or even his identity (e.g., his face may be captured in some video frames when the Amber Response is running while the vehicle is stopped and the user is walking in front of the camera). Performing in-vehicle image processing can minimize the need of transmitting the raw data to the external servers, minimizing the private information leakage. Among three implementations of Amber Response, from privacy perspective, the Cloud-based implementation is the worst, whereas the SmartCore-based has the minimum information leakage (it does not transmit the raw image at all and only shares the user's location when it detects a suspicious vehicle in the surroundings). The hybrid version, that only transmits plate areas and removes other objects in the environment from the image, also significantly enhances the privacy of the user (in comparison to the Cloud-based version). However, based on our empirical results, it might occasionally misdetect some other objects in the environment as plates and transmit some inessential images, that can be potentially processed to reveal the user's location (e.g., images of logos and flyers), along with images of interest (i.e., images that only contain plates of nearby vehicles).

\subsection{Application 2: Insurance Monitor}
\label{APP_2}
Usage-based (also referred to as Pay-As-You-Drive) insurance policies are envisioned as the future of auto insurance \cite{PRIPAYD}. Several insurance companies worldwide (for example, MetroMile \cite{METRO}) have already introduced new low-rate insurance plans for which they take traveling mileage, along with the driver's behaviors, into account. They currently collect the required information (for example, the vehicle's speed and odometer readings) from a dongle that plugs into the vehicle. Moreover, they commonly collect other types of data from the OBD port, including various diagnostic messages. 

Despite the potential benefits that insurance dongles have offered, their usage is currently limited due to privacy concerns and security threats. Gao et al. \cite{ELASTIC} have shown that vehicle's location can be easily tracked by processing the vehicle's speed data. Furthermore, security researches \cite {SEC_ATTACK_D2,FOSTER,SEC_ATTACK_D} have discussed common security vulnerabilities of third-party dongles and demonstrated real-world life-threatening security attacks. For example, Foster et al. \cite{FOSTER} have exploited security vulnerabilities of an insurance dongle (used by MetroMile \cite{METRO}) to send arbitrary unauthorized messages to the OBD port. They constructed an end-to-end security attack, highlighting the potential seriousness of existing security flaws. 

A few solutions have been discussed in the literature to address the privacy/security issues associated with the use of insurance dongles \cite{PRIPAYD,SOL2}. Such solutions commonly require \textit{a design change in the hardware (insurance dongle) or back-end infrastructures (insurance servers)}. They impose significant extra costs to companies due to at least one of the two following reasons. First, insurance companies already have \textit{millions of active dongles in the market} and changing the whole infrastructure (including dongles and servers) is very difficult (if not impossible). Second, to minimize design costs, insurance companies commonly use generic OBD dongles that are available from third-party companies, however, the proposed solutions require new dongles that are \textit{specifically designed for insurance companies}. Thus, insurance companies are unwilling to incorporate these solutions into their in-use scheme. 

Based on ProCMotive, we design and develop an application that enables security/privacy-friendly usage-based insurance, while imposing \textit{no design change (and consequently, no additional cost) to the insurance company}. On the user side, the proposed application utilizes the access control scheme offered by SmartCore to ensure that the dongle only performs its intended activities, preventing security attacks and minimizing the privacy leakage. Moreover, it uses the port management capability provided by SmartCore, along with data manipulation techniques, to remove inessential sensitive data from the raw data requested by the insurance dongle while maintaining the similar utility. 

\subsubsection{Implementation}
This application uses port management API offered by SmartCore, along with the access control scheme, to address the above-mentioned attacks: (i) it prevents security threats enabled by sending arbitrary messages from vulnerable insurance dongles (e.g., \cite{SEC_ATTACK_D,SEC_ATTACK_D2}), and (ii) address the attack against location privacy in which the insurance company can continuously track the user (e.g., \cite{ELASTIC}), as described next.

\noindent\textbf{Preventing security threats}: Previous research studies \cite {SEC_ATTACK_D,SEC_ATTACK_D2,FOSTER} have shed lights on one common security vulnerability of third-party dongles: dongles can be enforced (either remotely over the cellular network or within a short distance over Bluetooth connection) to send arbitrary messages to the OBD port. This vulnerability can potentially offer a direct access to several vital components and systems in the vehicle, enabling the attacker to perform various life-threatening attacks, ranging from remotely controlling the braking system \cite{FOSTER} to launching DoS attack against various built-in systems \cite{SECURITY_EXP}. 

Such attacks are feasible since OBD port, which has been originally designed for diagnosis, has two main limitations. First, it does not offer any security scheme to distinguish authorized messages from unauthorized ones, assuming that every OBD-connected dongle is trusted and is allowed to transmit all requests and access all components. Second, it utilizes a very simple congestion control protocol that always prioritize the messages with lower PIDs over others. This congestion protocol makes DoS attack against OBD easy: the attacker can only send packets with lowest possible PID (commonly, $PID=00$) to the OBD port at a high frequency \cite{SECURITY_EXP}. 

To address above-mentioned attacks, Insurance Monitor ensures that (i) the dongle can transmit a set of expected requests (i.e., data request that are essential for usage-based insurance) and (ii) the rate of request generated by the dongle always remains below a reasonable threshold. This threshold can be predetermined by examination of an insurance dongle in a trusted environment (based on our empirical results, for MetroMile dongle this threshold can be set to one request per second). 

Upon the attachment of the insurance dongle, using the Android application that we have developed to offer a user-friendly interface, the user can choose his insurance company. On SmartCore, the appropriate access control file will be automatically created (using access control API), and an upper bound will be set for the rate of requests that the dongle can generate (using port management API).\\
\noindent\textbf{Addressing the attack against location privacy}: Using port management API, the proposed application first captures packets from a third-party OBD dongle and forwards it to the vehicle. It then gets the response from the vehicle and modifies its data field using privacy-preserving functions in such a way that the insurance company can still get similar utility (e.g., can correctly compute the number of times the user has speeding violation). Eventually, it sends the modified response to the dongle. Different privacy-preserving functions can be used to minimize the information leakage, in particular, data shuffling, noise addition, and rounding techniques \cite{DATA_MAN1, DATA_MAN2} have been extensively discussed in the literature. Using these techniques, in the prototype implementation of Insurance Monitor, we have implemented three privacy preserving algorithms. 
\begin{enumerate}
\item Alg. 1: Shuffling: Given a window size $W$, this algorithms aggregates $W$ speed samples ($V=\{V_i, ..., V_w\}$) and returns a random permutation of them ($V^*$). 
\item Alg. 2: Rounding and then shuffling: Given a windows size $W$, for each sample of the vehicle's speed $V_i$, this algorithm first rounds $V_i$ (to the nearest integer, nearest five, or nearest ten), then aggregates and shuffles $W$ of them, and eventually returns the set $V^*$. 
\item Alg. 3: Noise addition: For each sample of the vehicle's speed $V_i$, this algorithm picks a float number $Z_i$ drawn from a uniform distribution with the range of $R_{uniform}$, i.e., $0<Z_i<R_{uniform}$, and returns $V^*_i=V_i+Z_i$.
\end{enumerate}
Next, we demonstrate how each of these algorithms enhance the privacy of the user and affect the utility. 

\subsubsection{Evaluation}
To verify that Insurance Monitor can address security and privacy attacks discussed earlier, we first implemented two known security attacks (the attack that enables an attacker to send arbitrary message \cite{FOSTER} and a DoS attack \cite{SECURITY_EXP}) and the privacy attack that uses speed data to track the user (Elastic Pathing \cite{ELASTIC}), and confirmed that both attacks work when Insurance Monitor is not active. We then examined if/how Insurance Monitor can address these attacks. We observed that when Insurance Monitor is active it only allows the Insurance dongle to read speed data and odometer data and actively blocks all other types of request. This limitation, which is imposed by Insurance Monitor on the dongle, completely prevents the first security attack, however, the attacker might still try to launch DoS by sending the allowed requests with a high frequency. We also observed that Insurance Monitor correctly regulates the rate of requests, i.e., it puts requests generated by the dongle in a queue and only transmits one request to the vehicle every second. This completely addresses the second security attack. 

In order to examine how effectively the privacy-preserving algorithms implemented in the prototype version of Insurance Monitor can address Elastic Pathing \cite{ELASTIC}, we examined how the accuracy of the attack, i.e., the distance between the estimated destination and the actual destination divided by the actual travelled distance reduces, and a speed-related utility required for the usage-based policy degrades when Insurance Monitor exploits privacy-preserving algorithms. The speed-related utility is defined as the number of times that the speed is above a certain threshold. In our experiments, we set the speed threshold to $25mph$, i.e., we assume that the insurance company intends to know how many times the vehicle's speed exceeded $25mph$ and we utilized the database provided in \cite{ELASTIC} that includes several streams of a vehicle's speed collected from real-world driving traces. 

\begin{figure*}[h]
\centering
\includegraphics[trim = 30mm 0mm 130mm 10mm,clip, scale=0.6]{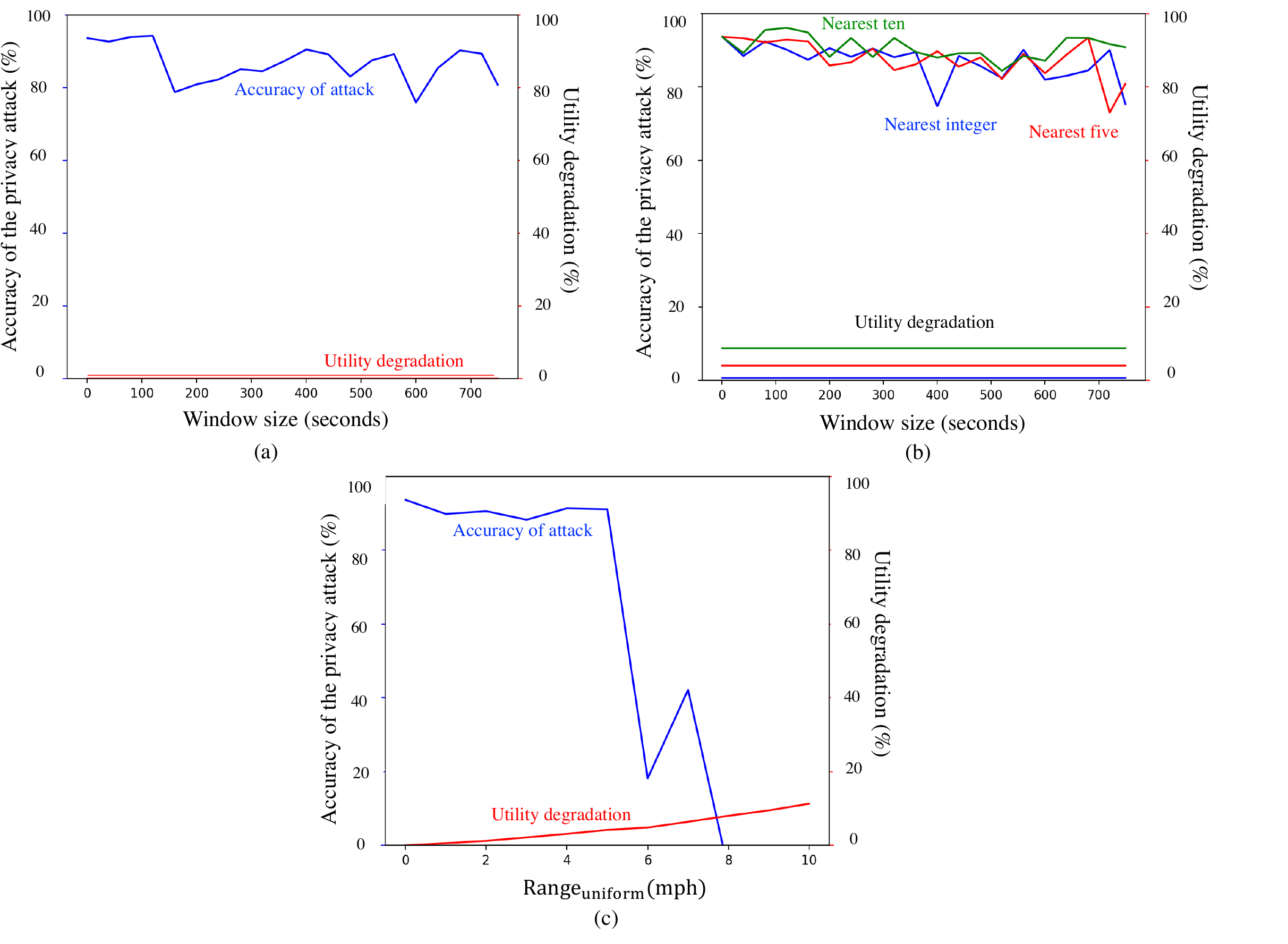}
\caption{Accuracy of the privacy attack (Elastic Pathing \cite{ELASTIC}) and utility degradation when privacy-enhancing algorithms  (Algs. 1,2, and 3) are used.}
\label{fig:ELASTIC_RESULS}
\end{figure*}

It is desired that Insurance Monitor reduces the accuracy of the attack, while maintaining the utility. Fig \ref{fig:ELASTIC_RESULS} shows both accuracy of the attack and utility degradation (i.e., the difference between computed utility based on the modified data and actual utility divided by the actual utility) for three algorithms discussed above. Fig \ref{fig:ELASTIC_RESULS} (a) demonstrates how windows size $W$ affects both utility and attack accuracy when Alg. 1 is used. Fig \ref{fig:ELASTIC_RESULS} (b) shows how both window size $W$ and rounding precision affect utility and attack accuracy when Alg. 2 is utilized. Eventually, Fig \ref{fig:ELASTIC_RESULS} (c) demonstrates the accuracy of attack and utility degradation with respect to the range of the uniform distribution ($R_{uniform}$) for Alg. 3. Based on our experimental results, Algs. 1 and 2 slightly decrease the accuracy of the attack (or equivalently, enhance the user's privacy), whereas Alg. 3 can significantly reduce the accuracy of the attack with minimal utility degradation.   

\subsubsection{Summary}
Table \ref{table:summary} highlights the advantages of ProCMotives for two above-mentioned vehicular applications. 

Comparing SmartCore-enabled (SmartCore-based and hybrid) implementations of Amber Response to its Cloud-based implementation (the baseline) demonstrates that they significantly reduce cellular data usage, enhance user privacy, and are more resilient against the potential unavailability of wireless connectivity while they can provide promising performance results. As demonstrated earlier, although we have utilized a very powerful Cloud server (with 8 CPUs and 32GBs of RAM), performance results provided by SmartCore-enabled implementations are comparable to the Cloud-based implementation for low frame rates. In particular, with video inputs captured at 1FPS, the SmartCore-based version of Amber Response can accurately (with the accuracy of $\%100$) detect the abductor's vehicle even faster than Cloud-based version (Table \ref{table:perf1200}). 

Furthermore, Insurance Monitor can provide several benefits for already-in-use usage-based insurance policies (the baseline). In particular, it can prevent several security attacks and significantly enhance the privacy of the vehicle's owner while maintaining the utility needed by insurance policies.

\begin{table*}[ht] 
\caption{Comparison of ProCMotive-enabled implementations with their baselines} % title of Table 
\centering % used for centering table 
\begin{tabular}{l c c c c c} % centered columns (4 columns)
\hline
Applications & Privacy & Security & Performance & Cellular data usage & Resiliency against \\
& & & & & connection unavailability \\
\hline\hline %inserts double horizontal lines 
Amber Res. (SmartCore) &  $\uparrow  \uparrow$  & N/A & Similar & $\downarrow  \downarrow$ & $\uparrow  \uparrow$ \\
 & & & at 1FPS & & \\
\hline
Amber Res. (Hybrid) & $\uparrow$ & N/A & Similar & $\downarrow$ & Similar \\
 & & & at 1FPS & & \\
\hline\hline\hline
Insurance Monitor & $\uparrow$ & $\uparrow$ & N/A & Similar & Similar \\
\hline %inserts single line 
\end{tabular} 
\label{table:summary}
\end{table*}

\section{Related work}
\label{RELATED}
The emergence of the IoT paradigm has led to an exponential increase in the number of Internet-connected sensing and computing objects, and in particular, has provided the opportunity to transform an isolated vehicle into an Internet-connected smart object \cite{IOT_SURVEY}. Several recent publications have discussed potential benefits that Cloud-based services can provide for Internet-connected vehicles and proposed novel architectures \cite{ARCH_C1,ARCH_C2,ARCH_C3} to enable Cloud-based services for \textit{future vehicles}. Furthermore, many researchers and developers have investigated novel Cloud-enabled vehicular applications \cite{APP_C1, APP_C2, APP_C3, APP_C4,APP_C5}. For example, Ji et al. \cite{APP_C1} have proposed a Cloud-based car parking system that aims to find the nearest available car parking lot by processing the data collected from nearby vehicles on the Cloud. Meseguer et al. \cite{APP_C5} have implemented a Cloud-based smartphone-assisted system that continuously analyzes drivers' behaviors using a neural network. 

In addition to Cloud-based services, different smartphone-based applications have been developed for diagnostic purposes \cite{APP_S1} (e.g., finding a faulty unit), controlling the vehicle's basic components \cite{APP_CONT1} (e.g., locking/unlocking doors), and assessing the driver's behaviors \cite{APP_S3}. 

Despite the existence of several proposal for development of vehicular applications in the literature, there is still a significant challenge that hinders their deployment: the majority of already-in-market vehicles have limited resources and communication capabilities and rarely support programability. Furthermore, as described earlier in Section \ref{SHORT_C}, mission-critical operations cannot be implemented on the Cloud or nearby personal devices due to connectivity and reliability issues. ProCMotive provides an interoperable approach to shift computational/storage resources from the Cloud to the vehicle, considering several shortcomings of previous approaches and different domain-specific design goals. Its unique approach imposes no design change on vehicles, and therefore, can potentially facilitate rapid development and deployment of new vehicular applications.

\section{Conclusion}
\label{CONC}
In this paper, we presented a reference architecture that potentially enables rapid development of various vehicular applications. The architecture is formed around a core component, called \textit{SmartCore}, a privacy/security-friendly programmable dongle that offers in-vehicle computational and storage resources and hosts applications. 

Based on the reference architecture, we developed an application development framework for vehicles, that we call \textit{ProCMotive}. To highlight potential benefits that ProCMotive offers, we proposed and developed two new vehicular applications based on ProCMotive, namely, Amber Response and Insurance Monitor. We evaluated these applications using real-world data and compared them with state-of-the-art technologies. 

ProCMotive enables application developers and researchers, who are interested in proposing and examining vehicular applications, to rapidly design, prototype, and evaluate novel applications for vehicles.  

\bibliographystyle{ACM-Reference-Format}
\bibliography{ProCMotive-Ref}

\end{document}